\documentclass[cameraready]{Interspeech}

\title{Neural Multichannel Distant Speaker Diarization and Source Separation\\with Beta Speaker Activity Prior}

\author[affiliation={1}, correspondingauthor]{Sicheng}{Mao}
\author[affiliation={1}]{Mathieu}{Fontaine}
\author[affiliation={2}]{Anthony}{Larcher}
\author[affiliation={1}]{Roland}{Badeau}

\address{
    $^1$ LTCI, Telecom Paris, Institut Polytechnique de Paris, France\\
    $^2$ LIUM, Universite du Mans, France
}

\email{prename.surname@telecom-paris.fr, anthony.larcher@univ-mans.fr}


\newcommand{\grey}[1]{\textcolor{gray}{#1}}
\usepackage{multirow, booktabs}
\usepackage[normalem]{ulem}
\usepackage{comment}
\useunder{\uline}{\ul}{}

\newcommand{\diag}{\mathrm{diag}}

\newcommand{\setRp}{\mathbb{R}_+}

\newcommand{\eye}{{\bf I}}

\newcommand{\distnormal}[2]{\mathcal{N}\left({#1}, {#2}\right)}
\newcommand{\distcmpnormal}[2]{\mathcal{N}_{\mathbb{C}}\left({#1}, {#2}\right)}

\NewDocumentCommand\newletter{m m o m m}{%
\NewDocumentCommand#1{s t@ o}{%
\IfBooleanTF{##1}{\mathbf{\MakeUppercase{#2}}\IfValueT{#3}{^{#3}}}{%
\IfBooleanTF{##2}{\mathbf{#2}\IfValueT{#3}{^{#3}}_{\IfValueTF{##3}{##3}{#5}}}{%
{#2}\IfValueT{#3}{^{#3}}_{\IfValueTF{##3}{##3}{#4}}%
}}}}

\NewDocumentCommand\newletterbm{m m o m m}{%
\NewDocumentCommand#1{s t@ o}{%
\IfBooleanTF{##1}{\bm{\MakeUppercase{#2}}\IfValueT{#3}{^{#3}}}{%
\IfBooleanTF{##2}{\bm{#2}\IfValueT{#3}{^{#3}}_{\IfValueTF{##3}{##3}{#5}}}{%
{#2}\IfValueT{#3}{^{#3}}_{\IfValueTF{##3}{##3}{#4}}%
}}}}

\newletter{\x}{x}{ftm}{ft}

\newletterbm{\psd}{\lambda}{nft}{ft}

\newcommand{\scm}[1][nf]{\mathbf{R}_{#1}}

\newcommand{\Q}[1][f]{\mathbf{Q}_{#1}}
\newletter{\q}{q}{fmm}{fm}
\newletter{\xt}{\tilde{x}}{ftm}{ft}
\newletter{\yt}{\tilde{y}}{nftm}{nft}

\newletter{\z}{z}{ntd}{nt}
\newletter{\zs}{z}[*]{ntd}{nt}
\newletter{\msk}{u}{nt}{t}

\newletter{\g}{w}{nfm}{nf}

\newcommand{\dec}[1][\theta,f]{g_{#1}}

\newletter{\src}{s}{nft}{nf}

\begin{document}

\maketitle

\begin{abstract}
Distant speaker diarization remains challenging due to adverse acoustic conditions, varying numbers of speakers and overlapping speech. While data-driven approaches have shown strong performance, model-driven methods offer a compelling alternative by leveraging spatial information from multichannel recordings.
This paper is motivated to propose a Bayesian diarization model for a model-driven method called neural FCASA to enhance its robustness.
Specifically, we propose a beta prior over speaker activity and hence a variational lower bound objective that can be seen as a regularized continuous speaker activity score in place of the original cross-entropy loss to train the diarization model.
Our experiments show significant improvements in terms of Diarization Error Rate by at least 3\% (16\% relatively) and Jaccard Error Rate by at least 4\% (20\% relatively) on the AMI dataset compared to the baseline.
\end{abstract}

\section{Introduction}
\label{sec:intro}

Speaker diarization (SD) addresses the problem of ``who spoke when'' by identifying and logging speaker-specific speech events within audio recordings. Distant speaker diarization remains a particularly challenging task due to difficult acoustic conditions encountered in settings such as meetings~\cite{pardoSpeakerDiarizationMultiple2006,watanabeCHiME6ChallengeTackling2020,cornellCHiME8DASRChallenge2024}, where speech is typically captured by distant microphone arrays and subject to noise and reverberation, a variable number of active speakers, and frequent speech overlaps.

To improve the robustness of distant diarization systems, data-driven approaches leverage data augmentation strategies incorporating noise, reverberation and synthesized overlapping speech to train a powerful speaker feature extractor~\cite{snyderDeepNeuralNetwork2017} in pipeline-based systems~\cite{landiniBayesianHMMClustering2022} or directly an end-to-end neural diarization (EEND) framework~\cite{fujitaEndEndNeuralSpeaker2019, horiguchiEncoderDecoderBasedAttractors2022}. This paradigm has recently been extended to Large Language Models (LLMs)~\cite{wangDiarizationLMSpeakerDiarization2024, yinSpeakerLMEndtoEndVersatile2025}, which demonstrate scaling laws on the diarization task.

In parallel, model-driven approaches retain considerable interest by integrating classical signal processing knowledge, particularly for multi-channel distant speech recordings. The direction-of-arrival (DOA), inter-channel time delay (ITD) and other distant speech-specific features captured by multi-channel recordings can be used to characterize the speech sources and hence contribute to the multi-channel diarization task. \cite{angueraAcousticBeamformingSpeaker2007, mariotteASoBOAttentiveBeamformer2024} exploit beamforming models and estimate the steering vectors to enhance speech representations. Another line of work proposes multichannel EEND~\cite{horiguchiMultiChannelEndtoEndNeural2022}, which adapts the Transformer encoder to capture spatio-temporal information. Neural FCASA~\cite{bandoNeuralBlindSource2024} proposes a hybrid stochastic deep learning model to jointly perform multichannel speaker separation and diarization.
\begin{figure}[t]
    \centering \includegraphics[width=\linewidth]{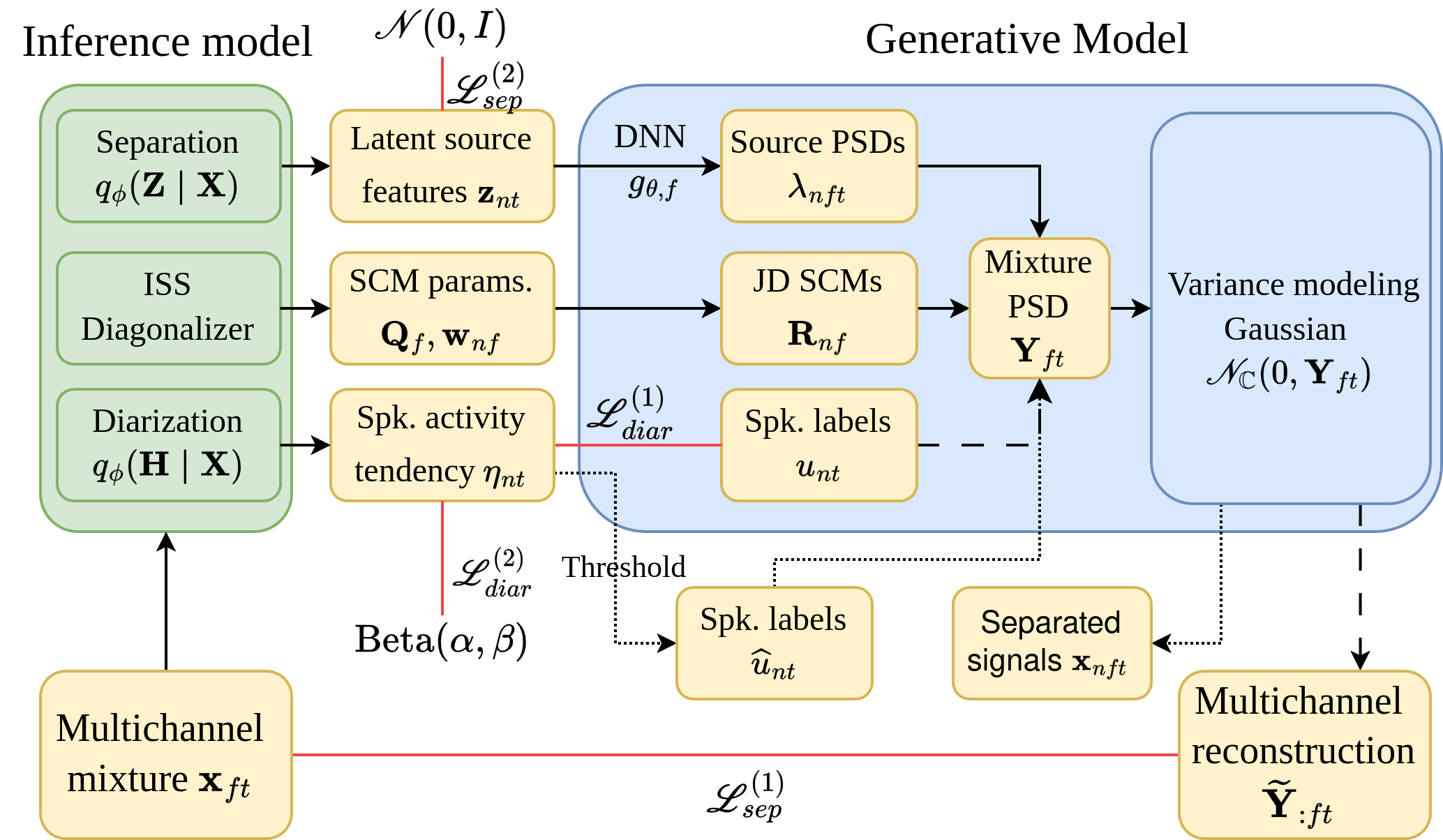}
    \caption{The overview of our model, training and inference}\small{Dashed: training pipeline; Dotted: inference pipeline; \\Solid: both; Red: loss terms.}
    \label{fig:beta-fcasa}
\end{figure}
More precisely, the neural FCASA focuses on source separation performance with the supervision of diarization labels. It proposes a Bayesian model for the source separation part of the joint learning framework, while leaving the diarization part learned in a non-Bayesian way.

In this paper, we focus on the diarization performance of the neural FCASA and improve it by proposing a generative diarization model that can be learned in a Bayesian manner as well. We introduce an explicit prior over the speaker activity variable using a beta distribution. This choice is motivated by several factors: (i) the beta distribution naturally models random variables defined on the unit interval, making it well-suited for speaker activity probabilities and (ii) it is conjugate to the Bernoulli likelihood, which yields a principled and tractable variational formulation.
Intuitively, this prior can be interpreted as modeling the internal state of a speaker that governs the tendency to speak or not at a given time step. It enables the incorporation of conversational dynamics into the diarization process, reflecting factors such as speaker behavior, interaction patterns, or conversational ambiance.  
Under this formulation, we manage to adopt a fully Bayesian approach for both the diarization part and the separation part of the model, and train the entire model using variational inference. 
We show that, under our modeling assumptions, the evidence lower bound (ELBO) of the diarization likelihood admits a closed-form expression.
We evaluate our approach on the AMI meeting corpus \cite{carlettaAMIMeetingCorpus2005} across multiple setups. Experimental results demonstrate significant improvements over the baseline and highlight the effectiveness of the proposed method.


\section{Background}
We describe the related baseline model neural FCASA, which can be seen as a variational autoencoder (VAE) with an encoder (aka. Inference model) and a decoder (aka. Generative model) as depicted in Figure \ref{fig:beta-fcasa}. We also provide insights about the beta distribution which will be chosen as a speaker activity prior. 
\subsection{Generative model in neural FCASA}
The neural FCASA proposes a multichannel speech mixture generation model as follows:
\begin{align}
  \z@ &\sim \distnormal{\bm{0}}{\eye},& \lambda_{nft} = \dec(\z@) \label{eq:deep-spec} \\
  \mathbf{Y}_{ft}&= \sum_{n=1}^N \msk \underbrace{\lambda_{nft} \scm}_{\triangleq \mathbf{Y}_{nft}},
  &\mathbf{x}_{ft} \sim \distcmpnormal{\bm{0}}{\mathbf{Y}_{ft}}. \label{eq:mixture}
\end{align}
where the deep spectral modeling~\cite{bandoNeuralFullRankSpatial2021} assumes the latent spectral characteristics $\mathbf{z}_{nt}\in \mathbb{R}^d$ 
follow the standard Gaussian distribution, and generate the power spectral density (PSD) $\lambda_{nft}\in \mathbb{R}_+$ of the source signal by a non-linear mapping $g_{\theta,f}$. The $N$ source PSDs are modulated by a source activity mask $\msk \in \{0,1\}$ and a spatial covariance matrix (SCM)~\cite{duongUnderDeterminedReverberantAudio2010,sawadaMultichannelExtensionsNonNegative2013} $\scm \in \mathbb{S}_+^{M\times M}$ 
and add up to the PSD $\mathbf{Y}_{ft}$ of the $M$-channel mixture signal $\mathbf{x}_{ft} \in \mathbb{C}^M$ in the Short Time Fourier Transform (STFT) domain, where $t=1,\ldots, T$ and $f=1,\ldots, F$ are the time and frequency indices respectively. 

\subsection{Inference model}
Given the observed mixture signals $\x*\triangleq \{\mathbf{x}_{ft}\}_{f,t=1}^{F,T}$ as input, an inference model $h_\phi$ is designed to estimate its latent source features $\z*\triangleq \{\z@\}_{n,t=1}^{N,T}$, speaker activity masks $\msk* \triangleq \{\msk\}_{n,t=1}^{N,T}$, and the spatial covariance matrices $\scm$. To this end, $h_\phi$ is constructed in a hybrid manner, which contains a neural encoder modeling the posterior distribution  $q_\phi(\z*|\x*), q_\phi(\msk*|\x*)$ of $\z*$ and $\msk*$, and an ISS diagonalizer~\cite{scheiblerSurrogateSourceModel2021} for computing $\scm$: 
\begin{align}
  \left\{ \Q[], \g*, q_\phi(\z*|\x*), q_\phi(\msk*|\x*) \right\} \leftarrow h_\phi(\x*), \label{eq:inference}
\end{align}
where the posterior distributions $q_\phi$ are defined by network outputs as:
\begin{align}
{q_{\phi}(\mathbf{Z}\mid\mathbf{X})\triangleq\prod_{n,t,d=1}^{N,T,D}\mathcal{N}\left(z_{ntd}~;~\mu_{\phi, ntd},\sigma_{\phi, ntd}^{2}\right),}\label{eq:infer-z}\\
{q_{\phi}(\mathbf{U}\mid\mathbf{X})\triangleq\prod_{n,t=1}^{N,T}\mathrm{Bernoulli}\left(u_{nt}~;~\eta_{\phi, nt}\right),}\label{eq:infer-u}
\end{align}
and the SCMs are further assumed to be jointly diagonalizable (JD) by a common projection matrix $\Q \in \mathbb{C}^{M\times M}$ across all the sources~\cite{sekiguchiFastMultichannelNonnegative2020,itoFastMNMFJointDiagonalization2019}, and $\g@ \in \setRp^M$ are the diagonal coefficients for source $n$, to facilitate the inference:
\begin{align}
  \scm = \Q^{-1} \diag(\g@) \Q^{-\mathsf{H}}. \label{eq:jd-scm}
\end{align}

Specifically, the inference model in Eq. (\ref{eq:infer-u}) for diarization is trained by a BCE loss with respect to the ground truth label while the inference model in Eq. (\ref{eq:infer-z}) for separation is trained in a Bayesian manner.

Once all the parameters are inferred from Eq.~\eqref{eq:inference}, the separation is done by estimating isolated image signals $\mathbf{x}_{nft}$ using a multichannel Wiener filter~\cite{bandoNeuralFastFullRank2023,sekiguchiFastMultichannelNonnegative2020} 
$\mathbf{x}_{nft}=\msk\mathbf{Y}_{nft}\mathbf{Y}_{ft}^{-1}\mathbf{x}_{ft}$ and the diarization is done by thresholding the logits $\eta_{\phi,nt}$ to 0 or 1. 



\subsection{The Beta distribution}
\label{sec:beta}
\begin{figure}
    \centering
    \includegraphics[width=\linewidth]{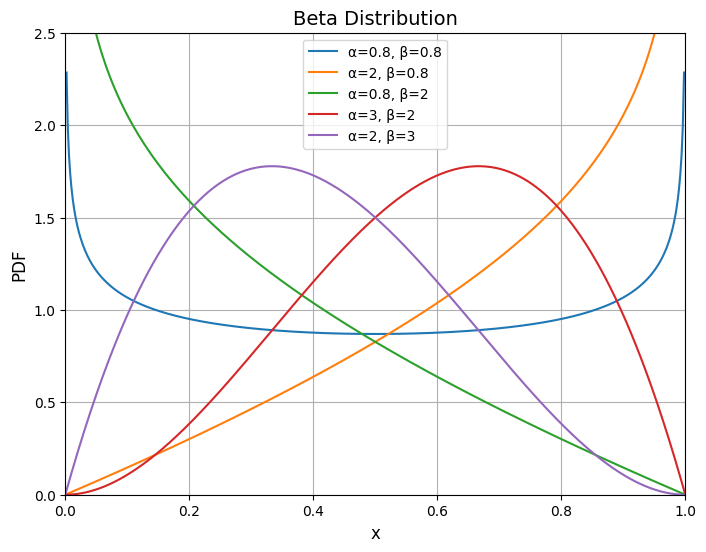}
    \caption{Beta Distribution with different hyperparameters}
    \label{fig:beta-dist}
\end{figure}

We recall some basic properties of the Beta distribution \cite{bishopPatternRecognitionMachine2006}. The density function of the Beta distribution is 
$$
\text{Beta}(x;\alpha,\beta)=\frac{\Gamma(\alpha+\beta)}{\Gamma(\alpha)\Gamma(\beta)}x^{\alpha-1}(1-x)^{\beta-1},
$$
where the support is $x\in (0,1)$, $\Gamma(\cdot)$ denotes the Gamma function, and the hyperparameters are $\alpha, \beta > 0$.
The shape of the distribution is determined by the values of the hyperparameters as depicted in Fig. \ref{fig:beta-dist}, which consists of three cases: (i) unimodal ($\alpha,\beta\geq1$), (ii) U-shaped/bimodal ($\alpha,\beta<1$) and (iii) monotonic ($\alpha\leq 1,\beta\geq1$ or $\alpha\geq1,\beta\leq1$).

\section{The Beta Neural FCASA}
\subsection{Generative modeling with beta speaker activity prior}
The inference model in Eq. (\ref{eq:infer-u}) for speaker activity in the neural FCASA involves two types of variables: 1) $u_{nt}$ denotes the observed speaker activity mask which is either $0$ (inactive) or $1$ (active); 2) $\eta_{\phi,nt}$ denotes the probability that speaker $n$ speaks at time $t$, which is continuous in $(0,1)$. We call the latter "speaker activity tendency", as it suggests the inner-state tendency of a speaker to speak, which may be determined by factors including but not restricted to 
\begin{itemize}
    \item the speaker's character, e.g. whether he/she is outgoing or shy to speak,
    \item the ambiance of the conversation, e.g. whether it is encouraging or discouraging for the attendees to speak.
\end{itemize}
Therefore, it is beneficial to explicitly model the speaker activity tendency with a prior to incorporate information of these factors. Specifically, since the speaker activity tendency is the parameter of a Bernoulli distribution, Bayesian statistics suggests its conjugate prior to model the distribution of its parameter, which is the beta distribution \cite{raiffa1961applied}.

The new generative model is hence as follows:
\begin{align}
    \eta_{nt} &\sim \text{Beta}(\alpha,\beta)\label{eq:beta}\\
    u_{nt} &\sim \text{Bernoulli}(\eta_{nt})\label{eq:bernoulli}
\end{align}
together with Eqs. (\ref{eq:deep-spec})-(\ref{eq:mixture}). We will refer to it as the generative modeling of the beta neural FCASA.

\subsection{The PERT parametrization}
\begin{figure}[t]
    \centering
    \includegraphics[width=\linewidth]{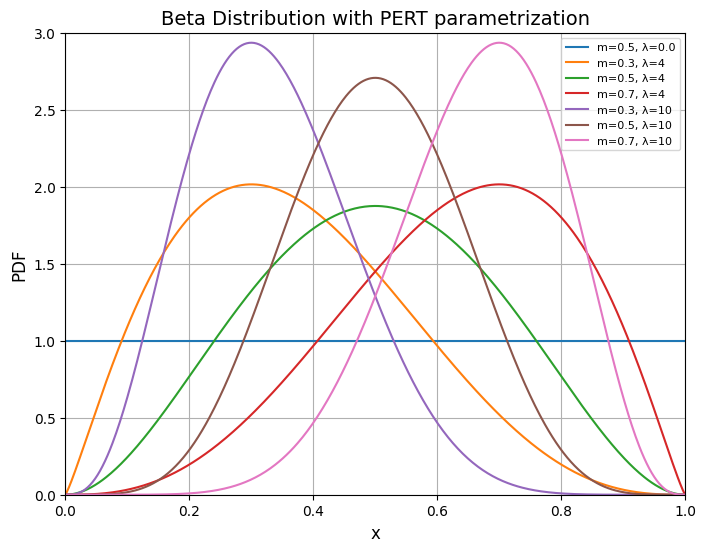}
    \caption{Beta distribution with different PERT hyperparameters}
    \label{fig:beta-pert}
\end{figure}

As we saw in Section \ref{sec:beta}, the canonical parametrization $\alpha,\beta$ does not have a decorrelated control over the shape of the beta distribution. To facilitate the learning of the hyperparameters, we adopt the PERT parametrization \cite{PERT}, which is a differentiable bijection when $\alpha,\beta\geq 1$:
\begin{align}
\mathrm{PERT}&:(\alpha,\beta) \mapsto \left(\frac{\alpha-1}{\alpha+\beta-2},\alpha+\beta-2\right),\\
\mathrm{PERT}^{-1}&: (m,\lambda) \mapsto (1 + \lambda m, 1 + \lambda ( 1 - m )),
\end{align}
where $m\in [0,1]$ and $\lambda \geq 0$ are the mode and the concentration of the beta distribution. Intuitively, the speaker activity tendency models the speaking will of a speaker, where $m$ represents the energy level to trigger speech and $\lambda$ represents how certain such decision is. Fig. \ref{fig:beta-pert} shows the shape change of the beta distribution according to the reparametrization of $m$ and $\lambda$.

To apply such parametrization, we restrict the paramters of the beta distribution to the case $\alpha,\beta\geq1$ in our model, which has a unimodal shape. We rule out the ``U-shape'' because the activity tendency for a typical speaker should be concentrated rather than dispersed — their inclination to speak or remain silent at any given moment is unlikely to be simultaneously strong in both directions. As to the monotonic case, it is included as a special case for the unimodal one when $\alpha=1, \beta\geq 1$, or $\alpha\geq 1, \beta=1$.

\subsection{The inference model}
With the explicit modeling of speaker activity tendency $\eta$, we adapt Eq. (\ref{eq:infer-u}) as follows:
\begin{equation}
{q_{\phi}(\mathbf{H}\mid\mathbf{X})\triangleq\prod_{n,t=1}^{N,T}\mathrm{Beta}\left(\eta_{nt}~\big|~\alpha_{\phi, nt},\beta_{\phi,nt}\right),}\label{eq:infer-eta}
\end{equation}
where $\mathbf{H}\triangleq \{\eta_{nt}\}_{n,t=1}^{N,T}$.
The $\alpha_{\phi,nt},\beta_{\phi,nt}$ are modeled by the outputs of the encoder of the neural FCASA model with the PERT parametrization:
\begin{align}
    m_{\phi,nt} &= \mathrm{sigmoid}(\text{Encoder}(h_m,h_{m\lambda})),\label{eq:encoder-m}\\
    \lambda_{\phi,nt} &= \mathrm{softplus}(\text{Encoder}(h_\lambda,h_{m\lambda})),\label{eq:encoder-lambda}\\
    (\alpha_{\phi,nt},\beta_{\phi,nt})&=\mathrm{PERT}^{-1}(m_{\phi,nt},\lambda_{\phi,nt}),
\end{align}
where $h_m, h_\lambda, h_{m\lambda}$ are hidden states that model $m,\lambda$ and the interdependency of $m,\lambda$. The sigmoid and softplus functions are applied to ensure the dynamic range of $m$ and $\lambda$.

\subsection{Training}
We train the parameters of the generative model in Eqs. (\ref{eq:deep-spec}), (\ref{eq:mixture}) and Eqs. (\ref{eq:beta}),  (\ref{eq:bernoulli}) by maximizing the log-likelihood $\log p_{\theta}(\x*)$ of the mixture observations $\x*$, which is similar to~\cite{bandoNeuralFastFullRank2023, bandoNeuralBlindSource2024}. We use the amortized variational inference to optimize the evidence lower bound~(ELBO), with the inference model in Eqs. (\ref{eq:infer-z}),(\ref{eq:infer-eta}) as the surrogate model. The ELBO simply majorizes the log-likelihood as follows:
\begin{align}
&\log p_{\theta}(\x*,\msk*)=\log\int p_{\theta}(\x*,\msk*,\mathbf{H},\z*)d\mathbf{H} d\z*\\
	&\geq\underbrace{\mathbb{E}_{q_{\phi}(\msk*)q_{\phi}(\z*)}(\log p_{\theta}(\x*\mid\msk*,\z*))}_{\mathcal{L}_{sep}^{(1)}}\underbrace{-\mathrm{KL}[q_{\phi}(\z*)\Vert p_{\theta}(\z*)]}_{\mathcal{L}_{sep}^{(2)}}\label{eqn:sep-loss}\\
    &+\underbrace{\mathbb{E}_{q(\mathbf{H})} [\log p(\msk*|\mathbf{H})]}_{\mathcal{L}_{diar}^{(1)}}\underbrace{-\text{KL}[q(\mathbf{H})\Vert p(\mathbf{H})]}_{\mathcal{L}_{diar}^{(2)}}\label{eqn:diar-loss},
\end{align}
where the separation loss $\mathcal{L}_{sep}$ remains the same as in the neural FCASA \cite{bandoNeuralBlindSource2024}. The diarization loss $\mathcal{L}_{diar}$ can be computed in closed form instead of using the Monte-Carlo approximation as in~\cite{kingmaAutoEncodingVariationalBayes2022}. Specifically, $\mathcal{L}_{diar}^{(1)}$ is
\begin{equation*}
\begin{aligned}
&E_{q(\eta)}[\log p(u|\eta)] 
 =\int_0^1 \log p(u|\eta) q(\eta) d\eta\\
 &=u\psi(\alpha_\phi)+(1-u)\psi(\beta_\phi)-\psi(\alpha_\phi+\beta_\phi)\\
\end{aligned}
\end{equation*}
where $\psi(\cdot)$ denotes the digamma function, and $\mathcal{L}_{diar}^{(2)}$ is the Kullback-Leiber (KL) divergence of two beta distributions \cite{rauberProbabilisticDistanceMeasures2008}:
\begin{equation*}
\begin{aligned}
&\text{KL}[q(\eta)\Vert p(\eta)] = \ln\frac{\Gamma(\alpha_\phi+\beta_\phi)\Gamma(\alpha)\Gamma(\beta)}{\Gamma(\alpha+\beta)\Gamma(\alpha_\phi)\Gamma(\beta_\phi)}\\
&+(\alpha_\phi-\alpha)\psi(\alpha_\phi)+(\beta_\phi-\beta)\psi(\beta_\phi)\\
&-(\alpha_\phi+\beta_\phi-\alpha-\beta)\psi(\alpha_\phi+\beta_\phi)\\
\end{aligned}
\end{equation*}
where $\alpha,\beta$ are predefined hyperparameters of the prior model in Eq. (\ref{eq:beta}).
Note that in the neural FCASA \cite{bandoNeuralBlindSource2024}, the log-likelihood of the diarization model is approximated by a cross-entropy loss and here we replace it by optimizing an ELBO function that supervises a continuous tendency variable with a KL divergence regularization.

In practice, the objective function is a weighted sum of these loss terms:
\begin{equation}
    \mathcal{L} = \mathcal{L}_{sep}^{(1)} + \gamma_1\mathcal{L}_{sep}^{(2)} +\gamma_2\mathcal{L}_{diar}^{(1)} +\gamma_3\mathcal{L}_{diar}^{(2)},\label{eq:objective}
\end{equation}
where $\gamma_1,\gamma_2,\gamma_3 >0 $ are hyperparameters.
\section{Experiments}
\label{sec:results}
To fairly compare with the neural FCASA as our baseline model, we follow the same experimental settings as in~\cite{bandoNeuralBlindSource2024}. 
\footnote{Codes available at \url{https://github.com/alephpi/neural-fcasa}.}
\subsection{Dataset}
\label{sec:dataset}
We evaluate the proposed method on the AMI corpus~\cite{carlettaAMIMeetingCorpus2005}.
This dataset contains about 100 hours of 16 kHz English meeting recordings with 3 to 5 participants.
Audio was captured using an 8-mic circular array with a radius of 10 cm placed on the table. Our experiments adopt the official split: training (80.7 h), development (9.7 h) and evaluation (9.1 h).

\subsection{Model architecture}
The network architecture of the beta neural FCASA follows that of the neural FCASA \cite{bandoNeuralBlindSource2024} 
with the modification mentioned in Eqs. (\ref{eq:encoder-m}), (\ref{eq:encoder-lambda}). Compared to the original model's 24 million parameters, such modification eventually adds only 257 parameters under the configurations mentioned below; therefore, it has a negligible impact on training and inference speed.
%
\subsection{Configurations}
\label{sec:config}
All signals are preprocessed with dereverberation using Weighted Prediction Error (WPE)~\cite{yoshiokaGeneralizationMultichannelLinear2012}. Spectrograms are computed via STFT with a window size of 512 and a hop size of 160. The model assumes $N=6$ sources, comprising 5 speaker channels ($n = 1,\ldots,N-1$) and one noise channel ($d = N$). The latent dimensionality is set to $d=10$ for the noise channel and $d=64$ for the speaker channels, in order to mitigate channel modeling ambiguity~\cite{bando22_interspeech}.
Speaker activations $\msk$ are derived from oracle diarization labels, while the noise channel remains permanently active to avoid conflation with silence. The weighting coefficients $\gamma_1,\gamma_2,\gamma_3$ in Eq.~\eqref{eq:objective} are all set to 1.0 by hypertuning on the development set.  

During training, recordings are split into 20-second segments, from which 10-second contiguous crops are randomly sampled and processed with a batch size of 128. The model is optimized over 200 epochs using the AdamW optimizer with a learning rate of $10^{-4}$ and weight decay of $10^{-5}$. A cyclic annealing schedule is employed to prevent KL vanishing. At inference, recordings are segmented into 10-second clips. The predicted speaker activity $\eta_{\phi,nt}$ is stabilized via median filtering over 11 frames and binarized using a threshold of 0.5.

Diarization performance is assessed using Pyannote~\cite{BredinPyannote2023}, reporting the Diarization Error Rate (DER) along with its components—miss (Miss), false alarm (FA), and speaker confusion (Conf.)—as well as the Jaccard Error Rate (JER).\footnote{Objective source separation metrics are not evaluated, as the AMI corpus comprises real conversations lacking isolated ground-truth references.} Four evaluation protocols are considered following~\cite{landiniBayesianHMMClustering2022}: \textit{Forgiving} applies a 0.25s collar excluding overlapping regions, \textit{Fair} applies a 0.25s collar including overlapping regions, and \textit{Full} applies no collar within overlapping regions. Additionally, an \textit{Overlap} protocol is adopted, which applies no collar but restricts evaluation to overlapping speech segments, targeting the principal challenge faced by diarization systems.

We train all our models on the AMI training set. We have trained the beta neural FCASA model with six different hyperparameter settings with $m=0.3,0.5,0.7$ and $\lambda=4,10$. We also retrained a baseline Gaussian model proposed in the neural FCASA.
For each setting, we keep the final checkpoint along with the other best checkpoints on the AMI development set, evaluate all the metrics and report the best one. 
\subsection{Results and discussions}
The evaluation results are illustrated in Table~\ref{tab:results}
\footnote{We omit the results of $\lambda=10$ as we found that the performance has no significant difference compared to that of $\lambda=4$ with the same $m$.}.
We first observe that for each model, the error rates become larger when the evaluation setup becomes stricter on overlaps. We then observe that every model has consistent improvements against the baseline in error rates across different setups. Specifically, the DERs have improved by 3\% to 4\% (by 16\% to 30\% relatively) and the JERs have improved by 4\% to 6\% (by 20\% to 27\% relatively) across different setups. The hyperparameter setting with $m=0.3, \lambda=4$ has the best performance in general, which suggests that the prior information in the conversational dynamics matters. Overall, the results justify the effectiveness of incorporating a speaker activity prior to the diarization model.


\begin{table}[t]
\resizebox{\columnwidth}{!}{
\begin{tabular}{clccccc}
\hline
$m,\lambda$ & \multicolumn{1}{c}{Setup} & Miss ($\downarrow$) & FA ($\downarrow$) & Conf. ($\downarrow$) & DER ($\downarrow$) & JER ($\downarrow$) \\ \hline
\multirow{4}{*}{(baseline)} &
  Forgiving &
  5.59 &
  8.16 &
  0.73 &
  14.48 &
  13.50 \\
 &
  Fair &
  8.61 &
  6.33 &
  0.79 &
  15.73 &
  23.16 \\
 &
  Full &
  10.77 &
  6.81 &
  1.16 &
  18.73 &
  27.65 \\
 &
  Overlap &
  19.40 &
  3.80 &
  0.90 &
  24.11 &
  26.53 \\ \hline
\multirow{4}{*}{0.3, 4} &
  Forgiving &
  5.37 &
  \textbf{4.31} &
  \textbf{0.43} &
  \textbf{10.11} &
  \textbf{9.73} \\
 &
  Fair &
  8.00 &
  \textbf{3.39} &
  \textbf{0.44} &
  \textbf{11.84} &
  \textbf{16.85} \\
 &
  Full &
  9.96 &
  \textbf{4.70} &
  \textbf{0.65} &
  {\ul 15.31} &
  \textbf{21.98} \\
 &
  Overlap &
  17.20 &
  \textbf{2.71} &
  \textbf{0.49} &
  {\ul 20.41} &
  \textbf{21.90} \\ \hline
\multirow{4}{*}{0.5, 4} &
  Forgiving &
  \textbf{4.55} &
  5.89 &
  0.50 &
  10.94 &
  {\ul 9.93} \\
 &
  Fair &
  \textbf{6.97} &
  4.69 &
  0.53 &
  12.19 &
  {\ul 17.04} \\
 &
  Full &
  \textbf{8.70} &
  6.07 &
  0.79 &
  15.55 &
  {\ul 22.21} \\
 &
  Overlap &
  \textbf{15.23} &
  3.47 &
  0.64 &
  \textbf{19.35} &
  {\ul 20.96} \\ \hline
\multirow{4}{*}{0.7, 4} &
  Forgiving &
  {\ul 5.16} &
  {\ul 4.64} &
  {\ul 0.49} &
  {\ul 10.29} &
  9.97 \\
 &
  Fair &
  {\ul 7.69} &
  {\ul 3.70} &
  {\ul 0.50} &
  {\ul 11.89} &
  17.40 \\
 &
  Full &
  {\ul 9.69} &
  {\ul 4.77} &
  {\ul 0.75} &
  \textbf{15.21} &
  22.57 \\
 &
  Overlap &
  {\ul 17.11} &
  {\ul 2.80} &
  {\ul 0.56} &
  20.48 &
  22.33 \\ \hline
\multicolumn{1}{l}{\begin{tabular}[c]{@{}l@{}}\grey{Pyannote 3.1}\\ \grey{IDM}\end{tabular}} &
  \grey{Full} &
  \grey{9.5} &
  \grey{3.6} &
  \grey{5.7} &
  \grey{18.8} &
  \grey{-} \\
\multicolumn{1}{l}{\begin{tabular}[c]{@{}l@{}}\grey{Pyannote 3.1}\\ \grey{SDM}\end{tabular}} &
  \grey{Full} &
  \grey{11.2} &
  \grey{3.8} &
  \grey{7.5} &
  \grey{22.4} &
  \grey{-} \\ \hline
\end{tabular}
}
\caption{Diarization metrics of the trained models on AMI evaluation set. All metrics are reported in percentage with the best score in \textbf{bold} and the second-best score {\ul underlined}. We also list the single channel diarization baseline Pyannote 3.1 \cite{BredinPyannote2023} as a reference. Note that the evaluation of our method is performed chunk-wisely (10 seconds), which is not directly comparable to the Pyannote baselines.}
\label{tab:results}
\end{table}

\section{Conclusion}
In this paper we improve the neural FCASA, a joint separation and diarization system, by incorporating an explicit prior model on the speaker activity based on the beta distribution.
By doing so, we formalize the training objective of the system in a full Bayesian manner and solve it with variational Bayes inference. We find that the ELBO of the diarization loss can be computed in closed form under our model assumptions and we use the PERT parametrization to facilitate parameter learning. Our experiments show significant improvements of speaker diarization performance in terms of Diarization Error Rate at least by 3\% (16\% relatively) and Jaccard Error Rate at least by 4\% (20\% relatively) compared to the baseline. Future work may focus on: (i) the estimation of the hyperparameter of the beta prior; (ii) a better conversation model for speaker activity that incorporates more complex conversational dynamics; and (iii) improving diarization with a better separation model.

\ifcameraready
\section{Acknowledgments}
We thank Dr. Yoshiaki Bando for insightful discussions. 
This project is funded by ANR Project SAROUMANE (ANR-22-CE23-0011) and granted access to the HPC resources of IDRIS under the allocation 20191014876 attribution made by GENCI.
\else
\fi

\section{Generative AI Use Disclosure}
We acknowledge the use of Claude (https://claude.ai/) to polish this paper. We pasted the draft of the abstract, the first three paragraphs of Section \ref{sec:intro}, 
Section \ref{sec:dataset} and Section \ref{sec:config} with the instruction ``Please rephrase the text in an academic manner''. The outputs were then modified further to better represent our own writing.




\bibliographystyle{IEEEtran}
\bibliography{abbr,mybib}

\end{document}